\documentclass[aip,
 amsmath,amssymb,
preprint,
]{revtex4-1}

\usepackage{xcolor}

\usepackage[russian, english]{babel}  
\usepackage[T1, T2A]{fontenc}  
\usepackage[utf8]{inputenc}   
\renewcommand{\figurename}{Figure}
\renewcommand{\tablename}{Table}

\usepackage{graphicx}
    \graphicspath{{figures/}} 
\usepackage{dcolumn}
\usepackage{bm}

\usepackage{mathptmx}
\usepackage{etoolbox}
\usepackage{mathtools,amssymb}

\usepackage{subcaption}  
\usepackage{caption}      

\usepackage{setspace}
\usepackage{scalerel}

\makeatletter
\def\@email#1#2{%
 \endgroup
 \patchcmd{\titleblock@produce}
  {\frontmatter@RRAPformat}
  {\frontmatter@RRAPformat{\produce@RRAP{*#1\href{mailto:#2}{#2}}}\frontmatter@RRAPformat}
  {}{}
}%
\makeatother

\begin{document}


\title[Energetics and limitations of passive electron transpiration cooling for hypersonic leading edges]{Energetics and limitations of electron transpiration cooling for hypersonic leading edges}
\author{B. Boyer}
\author{T.~S. Fisher}%
    \email{tsfisher@ucla.edu}
\affiliation{Mechanical and Aerospace Engineering Department, University of California Los Angeles, Los Angeles, California 90095 USA}

\date{\today}

\begin{abstract}
Electron transpiration cooling (ETC) offers a promising approach for thermal management of hypersonic vehicles by leveraging thermionic emission from the leading edge. While emitted electrons cool the surface, subsequent collection of flowfield electrons induces heating, limiting ETC effectiveness unless collection occurs in cooler aftbody regions. Most existing ETC studies neglect this heating contribution, assuming ideal downstream collection. This work integrates a one-dimensional collisionless plasma sheath model into a discretized leading edge framework to predict surface potentials and charged-particle fluxes. A parametric study examines how plasma and vehicle properties affect ETC performance. Results reveal that passive ETC is susceptible to thermionic space-charge overcompensation, which can reverse the intended cooling effect and cause surface heating at high plasma densities. Flowfield electron heating can be mitigated—but not eliminated—by using dielectric coatings or blunt geometries. The same mechanisms that protect blunt bodies from adverse electrical conduction and flowfield heating also preclude incorporation of ETC-based cooling in those vehicle geometries.
\end{abstract}

\maketitle

\section{Introduction}\label{sec:Introduction}

Electron transpiration cooling (ETC) has received attention in recent years as a theoretical thermal protection system (TPS) for hypersonic leading edges. ETC relies on thermal energy to free electrons from the surface through thermionic emission, which provides net cooling when emitted electrons are replaced by conduction electrons at lower energies. Researchers at BTSU Voenmeh considered thermionic cooling as an internal mechanism within a double hull made of thermionic cells \cite{kolychev2012active, kolychev2013active, kolychevparameters}, but ETC in an external configuration (and the acronym ETC itself) was first proposed by researchers at Lockheed Martin \cite{uribarri2015electron} and the University of Michigan \cite{alkandry2014conceptual}. Uribarri and Allen \cite{uribarri2015electron} compared ETC to the Stefan-Boltzmann Law and found that thermionic cooling would exceed blackbody radiative cooling for a surface with a work function of 2.8 eV at 2000 K, which is the temperature commonly associated with sustained hypersonic flight. Researchers at the University of Michigan incorporated ETC equations as boundary conditions in a CFD model, and predicted a 40\text{\%} reduction in the peak temperature for a wedge with a 1 cm leading edge radius and 2 eV work function flying at 6 km/s at 60 km altitude \cite{alkandry2014conceptual}.

These initial studies assumed ideal electron emission, but ETC performance can be limited by the formation of a retarding space-charge field in the sheath that inhibits emission when the emitted flux exceeds a critical value. Most ETC literature to date has focused on identification and mitigation of these space-charge limitations. The critical emission limit is proportional to the ion number density of the adjacent plasma, as the plasma sheath is positively charged and compensates for thermionic space-charge. If the plasma density is sufficientyly large, overcompensation of thermionic-space charge induces a strong electric field at the wall that can significantly enhance thermionic emission. ETC is therefore best-suited for extreme trajectories characterized by higher ionization fractions. It has been proposed that blunt reentry vehicles may attain substantial ETC cooling due to the high plasma density in the shock \cite{hanquist2018effectiveness, hanquist2019plasma}. For sharper leading edges at more moderate hypersonic trajectories, transpiration of readily ionized species (plasma seeding) has been proposed as a means to improve ETC performance \cite{tropina2020modeling, sahu2021sheath, andrienko2021computational, sahu2022plasma, sahu2022cesium, sahu2022full}.

\begin{figure*} [htb]
  \centering
  \begin{minipage}{0.48\textwidth}
    \includegraphics[width=\linewidth]{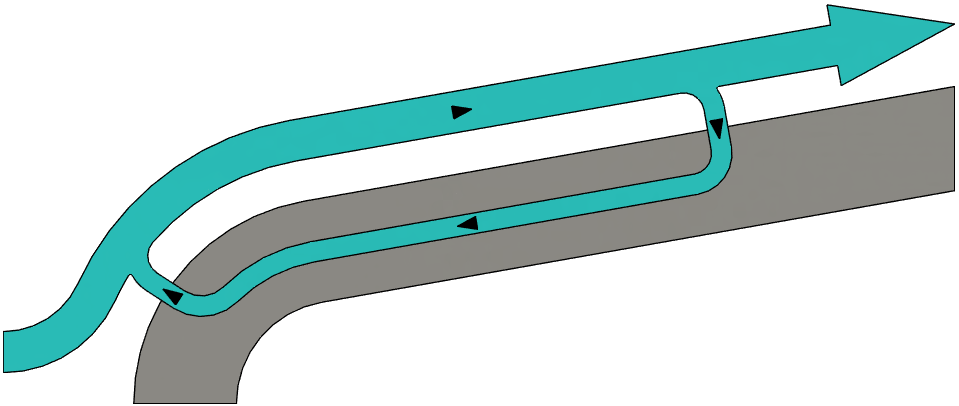}
    \caption*{a) Passive ETC}  
  \end{minipage}%
  \hspace{0.02\textwidth}
  \begin{minipage}{0.48\textwidth}
    \includegraphics[width=\linewidth]{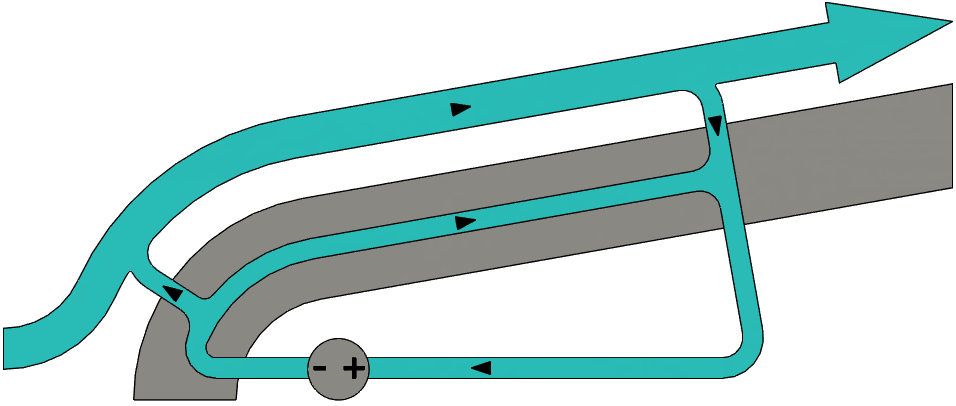}
    \caption*{b) Active ETC} 
  \end{minipage}
  \caption[Passive vs.~active ETC]{ETC circuits can utilize a) passive or b) active biasing of the leading edge. Passive ETC systems conduct electrons from colder aftbody regions towards the stagnation point. Active systems are likely to reverse the flow of electron conduction through the vehicle structure due to the applied potential difference.}
  \label{Fig: Active vs passive ETC}
\end{figure*}

However, most existing ETC literature has not fully accounted for plasma-wall interactions. Just as emitted electrons cool the surface, recaptured electrons locally heat the surface. Flowfield electron heating has generally been neglected due to an assumption that all emitted electrons are replenished by electron collection in designated aftbody regions where this heat can be more effectively managed. While such assumptions may be appropriate for active ETC systems with sufficient external bias, passive configurations (Fig.~\ref{Fig: Active vs passive ETC}) -- particularly those subjected to high plasma densities -- warrant closer scrutiny. The significance of ETC as a circuit was identified at its conception as a potential hypersonic TPS \cite{uribarri2015electron}, but has generally been neglected. Recent work has shifted focus back to the ETC circuit as a whole, with Monroe and Boyd \cite{monroe2025circuit} studying ETC circuits using a two-electrode configuration assuming ideal internal conductivity (i.e., no voltage drop within the electrodes), and Campanel et al.~\cite{campanell2025two} identifying an additional ``backflow saturation'' condition that can further limit thermionic current in a plasma diode system.  

This work proceeds in several parts.
Section~\ref{Sec: Basic theory} presents basic ETC theory, highlighting how overcompensation of thermionic space-charge can induce adverse thermal responses in passive configurations.
Section~\ref{Sec: ETC energetics} examines the energetics associated with charged-particle flows in ETC systems and derives an updated expression for the resulting heat flux.
A one-dimensional plasma sheath model is introduced in Section~\ref{Subsec: Plasma Model} to determine charged particle fluxes based on fixed wall and plasma properties.
This sheath model is incorporated into a discretized leading edge model in Section~\ref{Subsec:leading edge model} to compute self-consistent distributions of potential and particle flux along the surface.
Finally, Section~\ref{Sec: Cold ion Results} presents results from a parametric study assessing the influence of various plasma and wall properties on ETC performance.

\section{ETC Theory} \label{Sec: Basic theory}

The nominal thermionic emission flux is calculated using the Richardson-Dushman equation \cite{richardson1921emission}:
\begin{equation} \label{Eqn: Richardson-Dushman}
    J_t=J_R=\frac{A_R^*}{q}T_w^2\exp\left[{\frac{-q\phi_{WF}}{k_BT_w}}\right]
\end{equation}
$\phi_{WF}$ and $A_R^*$ are the work function and Richardson's constant for the surface. $k_B$, $T_w$, and $q$ are the Boltzmann constant, wall temperature, and electron charge magnitude respectively. Each emitted electron is assumed to cool the surface by an amount proportional to the work function plus $2k_BT_w$, the average kinetic energy of emitted electrons\cite{richardson1903xiii}:
\begin{equation} \label{Eqn: Basic thermionic electron cooling}
    Q_{ETC} = J_t(q\phi_{WF}+2k_BT_w)
\end{equation}

\begin{figure} [htb]
    \includegraphics[width=0.45\textwidth]{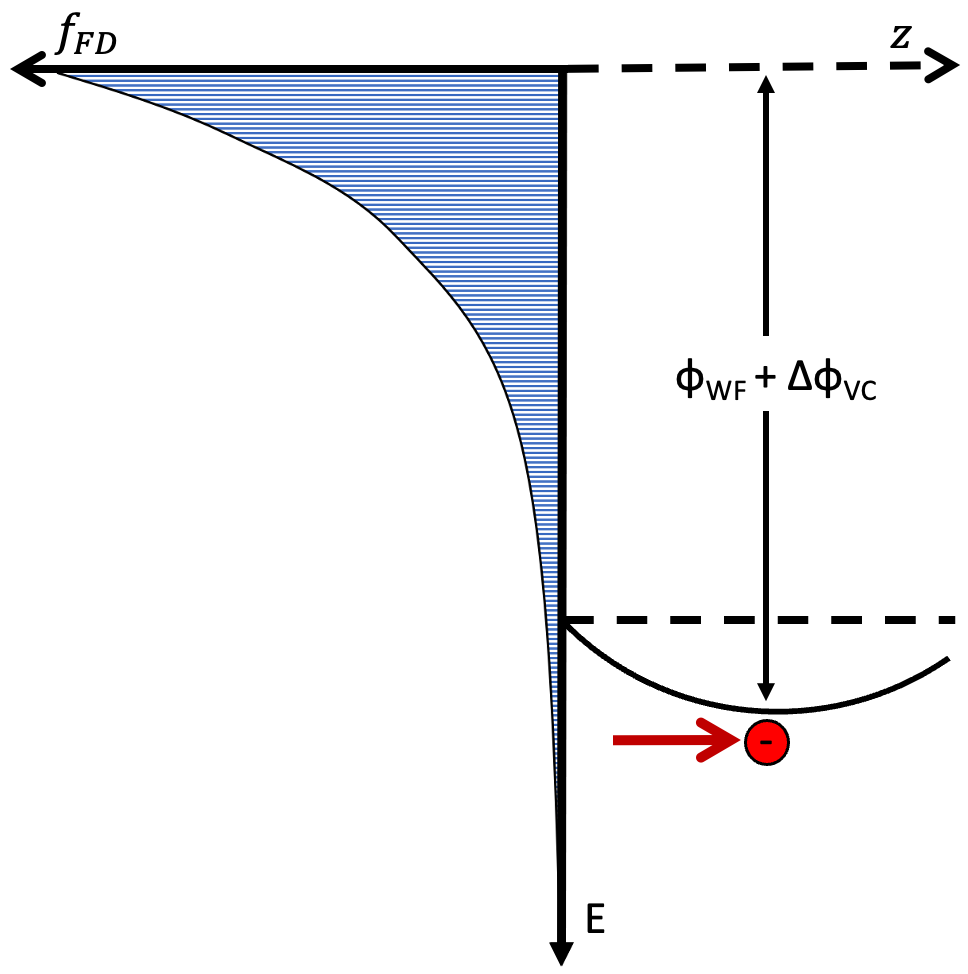}
    \centering
    \caption[Effect of space-charge on thermionic electron energy distribution]{Electron energies within the emitter material follow a Fermi-Dirac distribution, which is visualized on the left side of the figure. The population of electrons capable of emission from the surface is determined by truncating the Fermi-Dirac distribution such that the energy floor is equal to $\phi_{WF}$. When a virtual cathode forms near the surface, the emitted electron distribution is further truncated to a minimum energy of $\phi_{WF} + \Delta\phi_{VC}$.}
    \label{fig:Space charge visualized}
\end{figure}

The derivation of the Richardson current assumes ideal emission, where the thermionic flux is governed solely by properties of the wall. However, thermionic current can be limited by the formation of a retarding space-charge field that inhibits emission when the emitted flux exceeds a critical value (Fig. \ref{fig:Space charge visualized}).
When a quasineutral plasma encounters the emitter, electrons recombine at a faster rate than ions due to the difference in their thermal velocities. This phenomenon creates a thin sheath region where the plasma is positively charged relative to the wall \cite{francis2016introduction}. The ions within this region act to compensate for the large thermionic electron densities near the wall and increase the permitted current before space-charge limits are reached. 

\begin{figure}
    \includegraphics[width=0.4\textwidth]{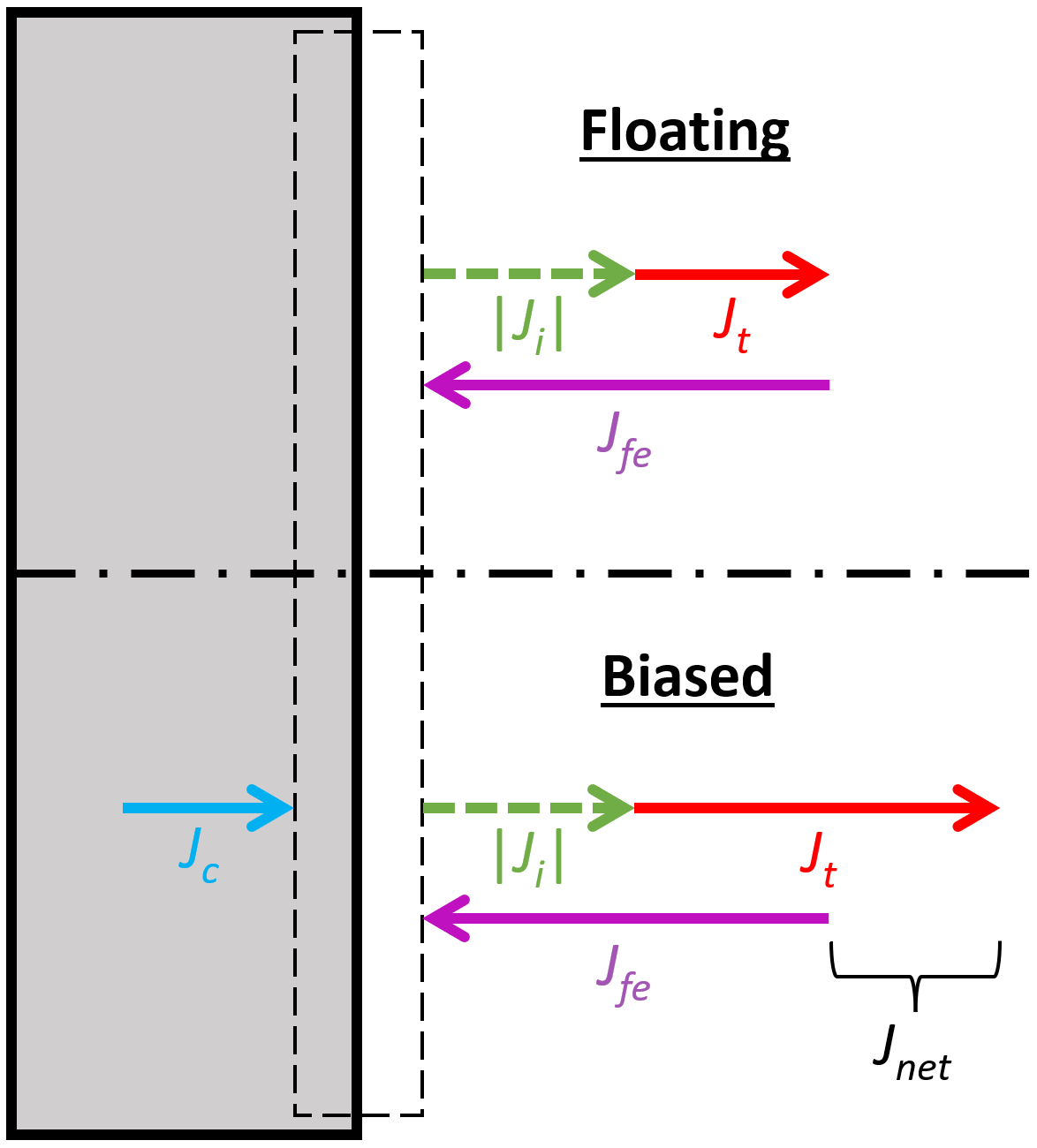}
    \centering
    \caption[Floating vs.~biased emission]{A floating surface allows zero net current between the wall and the plasma, while a biased surface permits net emission (or collection) of electrons from the wall to the plasma, which are then replaced through passive or active biasing of the surface.}
    \label{fig:Floating vs biased}
\end{figure}

The critical emission current for plasma adjacent surfaces for ETC applications is generally calculated using one of two limiting-case assumptions: biased and floating emission (Fig. \ref{fig:Floating vs biased}).
Floating emitters assume that the emitted electron flux equals the difference between the incoming flowfield electron and ion fluxes such that the net local current is zero:
\begin{equation} \label{Eqn: floating condition}
    J_t=J_{i}-J_{fe} 
\end{equation}
where the subscripts $i$, $fe$, and $t$ denote that a property corresponds to ions, flowfield electrons, and thermionic electrons, respectively. The wall repels flowfield electrons, and their flux, $J_{fe}$, is a function of the wall potential relative to the sheath edge. As ions are accelerated towards the wall, mass conservation dictates that the ion flux, $J_{i}$, is constant at all locations within the collisionless sheath and independent of the wall potential \cite{riemann1991bohm}:
\begin{gather}
    J_{fe} = n_{fe0}\sqrt{\frac{k_{B}T_{fe0}}{2\pi m_{e}}}\exp{\left(\frac{q\phi_{w}}{k_{B}T_{fe0}}\right)} \label{eqn: flowfield electron flux}\\
    J_{i} = n_{i0}u_B \label{eqn: ion flux}
\end{gather}
The subscript $0$ indicates that the property is taken at the sheath edge, and $n$ and $T$ are the number density and temperature. $n_{i0}$ is also called the plasma density and is equal to $n_{fe0}$ for non-emitting walls. $u_B$ is the Bohm velocity, which describes the ion velocity at the sheath edge after acceleration through the presheath region \cite{riemann1991bohm}. 
Increasing $\phi_w$ (i.e., making $\phi_w$ less negative) increases the electron flux to the wall while keeping the ion flux constant, thereby enhancing the allowable emission current per Eq.~\ref{Eqn: floating condition}. This trend continues until a critical potential is reached where the electron density near the wall is large enough relative to the ion population that a virtual cathode forms. This condition corresponds to the maximum emission current from a floating electrode exposed to a plasma at a given temperature and density, for which Hobbs and Wesson derived the following expression assuming cold emission and cold ions \cite{hobbs1967heat}:
\begin{gather}
    J_{crit, floating} = \left[\frac{C_{crit}}{1-C_{crit}}\right]J_i \\
    C_{crit}=1-8.3\sqrt{\frac{m_e}{m_i}}
\end{gather}

Biased systems assume that emitted electrons are replaced by a passive or external biasing of the cathode, permitting a net current through the plasma sheath. Eqs.~\ref{eqn: flowfield electron flux} and \ref{eqn: ion flux} still apply, but the wall potential is no longer constrained by the thermionic current and Eq.~\ref{Eqn: floating condition}. Most ETC research for biased wall conditions has relied on analytical expressions derived by Takamura et al.~that predict the critical emission current for cold emission \cite{takamura2004space}:
\begin{equation} \label{Equation: Takamura Warm}
    J_{crit, biased} = \frac{G\sqrt{-\pi \eta_{w}}}{1+G}\sqrt{\frac{2m_{i}}{\pi m_{e}}} J_{i}
\end{equation}
where G is a function of $\eta_{w}$, which is the wall potential normalized by $k_BT_p/q$. Hanquist et al.~later derived an expression similar to that of Takamura, which also accounts for the kinetic effects of the flowfield plasma, is easier to implement into CFD simulations, and shows better agreement with a one-dimensional direct kinetic simulation \cite{hara2018test}.

\begin{figure}
    \includegraphics[width=0.75\textwidth]{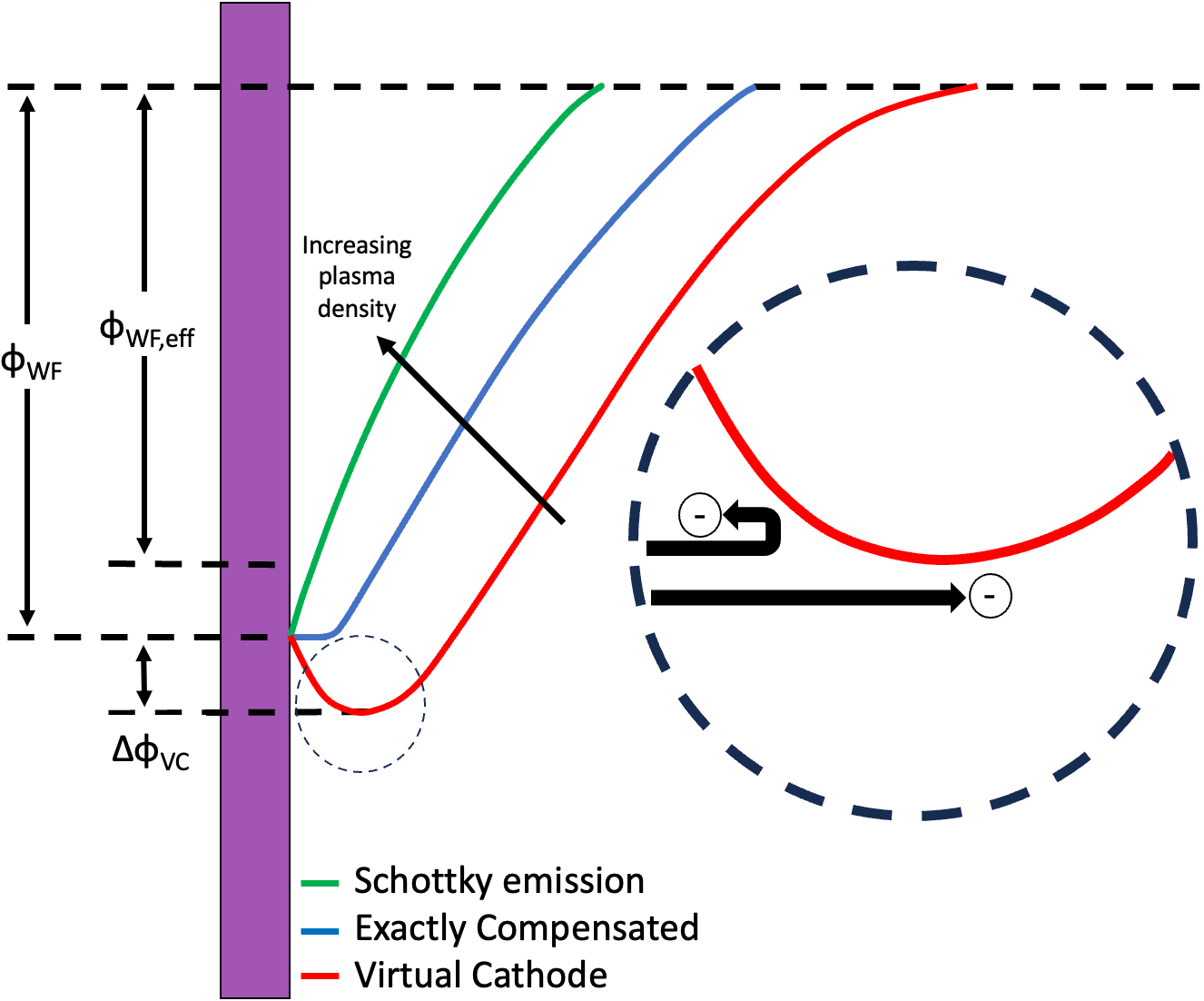}
    \centering
    \caption[Plasma compensation of thermionic space charge]{A plasma sheath is shown adjacent to an emitting wall for varying plasma densities while the wall potential and emission properties are held constant. When the flowfield plasma density is small relative to the thermionic current, a virtual cathode forms that reflects lower energy thermionic electrons to the surface. As the plasma density increases, the virtual cathode becomes weaker and eventually disappears when the electric field at the wall is zero. At this condition, the thermionic space-charge is said to be exactly compensated. Further increases to the flowfield plasma density result in an electric field that enhances electron emission through Schottky enhancement.}
    \label{fig: Plasma compensation}
\end{figure}

A floating emitter is more restrictive than a biased emitter due to the additional constraint imposed by Eq.~\ref{Eqn: floating condition}. However, the critical flux in both cases scales linearly with plasma density through the ion flux at the wall given by Eq.~\ref{eqn: ion flux}. Even a floating emitter can permit the full Richardson-Dushman flux if the plasma density is sufficiently large. If the plasma density further increases, the thermionic space-charge becomes overcompensated and the emission current increases through Schottky enhancement, in which the electric field at the wall decreases the effective work function for emitted electrons \cite{schottky1914einfluss, orloff2017handbook}: 
\begin{equation} \label{Eqn: Schottky}
    \phi_{WF,eff}=\phi_{WF}-\frac{1}{q}\sqrt{\frac{q^3\lvert E_w\rvert}{4\pi\epsilon}}
\end{equation}
where $\varepsilon$ is the permittivity of free space. The impact of increasing plasma density on the sheath structure, and the associated change from space-charge-limited emission to an overcompensated Schottky-enhanced regime, is shown in Fig. \ref{fig: Plasma compensation}.
It has been proposed that blunt reentry vehicles may attain substantial ETC cooling due to the high plasma density in the shock, despite likely functioning as floating emitters \cite{hanquist2018effectiveness, hanquist2019plasma}. For non-reentry applications, ETC is best-suited for extreme trajectories that induce higher ionization fractions. Transpiration of readily ionized species (plasma seeding) has been proposed for air-breathing vehicles at more moderate hypersonic trajectories to improve ETC performance \cite{tropina2020modeling, sahu2021sheath, andrienko2021computational, sahu2022plasma, sahu2022cesium, sahu2022full}. 

Consider the ideal operation of an ETC-based TPS. Cooling is achieved at the leading edge when high-energy electrons escape from the wall by overcoming the material work function, $\phi_{WF}$, at the surface. Charge conservation dictates that during steady state operation all emitted electrons must be reabsorbed somewhere along the vehicle surface. Just as emitted electrons cool the surface, recaptured electrons locally heat the surface by an amount proportional to the work function. It is readily apparent that any practical ETC TPS would recover a significant portion of the emitted electrons in colder regions of the aircraft, where that heat can be more effectively managed. Practical ETC must therefore form a circuit that recirculates collected electrons from aftbody regions to the leading edge.

This assumption directly contradicts those for a floating surface. By definition, a floating surface has zero net emission current, and any cooling associated with the work function is at best counteracted by an equivalent heating from flowfield electrons. As the plasma is hotter than the wall, each thermionic flowfield pair results in a slight heating of the surface. Ion bombardment further increases surface heating, and it becomes clear that ETC cooling is impossible regardless of thermionic flux if a hypersonic vehicle behaves as a floating surface. It is readily apparent that the origin of replacement electrons is critical to accurately predict ETC performance.

\begin{figure*}
    \centering
    \includegraphics[width=0.7\textwidth]{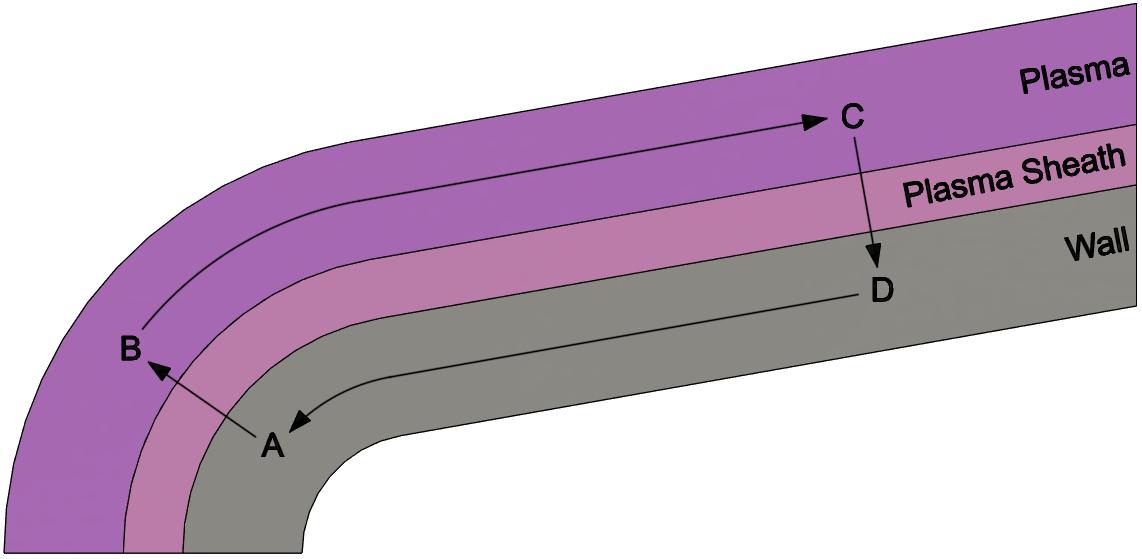}

    \vspace{0.5cm} 

    \includegraphics[width=0.9\textwidth]{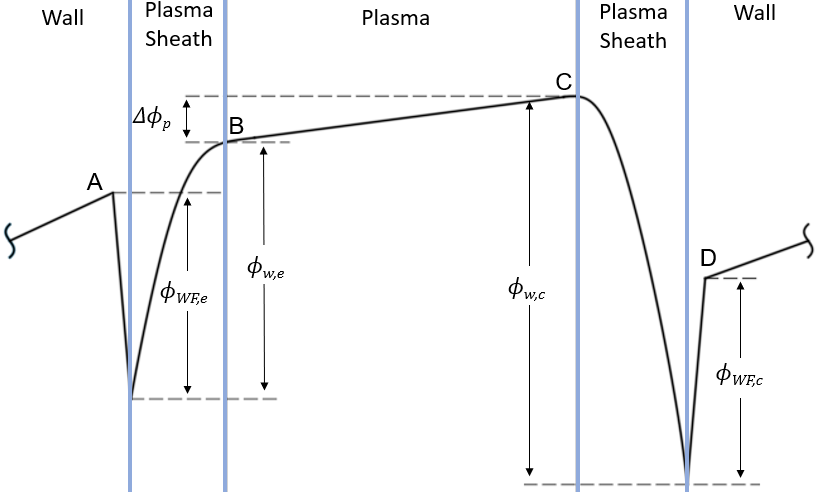}

    \caption[Potential diagram of ETC circuit]{Diagram of the ETC circuit with corresponding potential diagram. Regions are separated into wall, plasma sheath, and bulk plasma regions, whose sizes have been adjusted for visual clarity. Though the circuit is drawn using an arbitrary pair of emitter and collector points, charge exchange between the wall and plasma must be considered for all locations along the vehicle with a conductive surface.}
    \label{Fig: Full ETC circuit}
\end{figure*}

Suppose that a passive ETC TPS is undergoing ideal operation and all emitted electrons are reabsorbed downstream in a designated collector region. Internal electron conduction towards the leading edge results in a potential drop along the vehicle surface per Ohm's law. Conversely, the bulk plasma potential \emph{increases} along the vehicle surface due to thermionic electron conduction downstream within the plasma. A potential diagram between a single emission and collection point is shown in Fig.~\ref{Fig: Full ETC circuit}. As the wall potential relative to the sheath edge shields the surface from flowfield electrons and governs the flux through Eq.~\ref {eqn: flowfield electron flux}, the diverging potential lines indicate an increase in the kinetic energy requirement for absorption as the distance from the leading edge increases.

Electron collection is not a problem for conventional thermionic energy converters (TEC), which are characterized by thin, collisionless electrode gaps. Again using Fig.~\ref{Fig: Full ETC circuit}, the net thermionic current from the cathode through a vacuum gap is simply approximated by adding the potential difference between the two surfaces to the work function in Eq.~\ref{Eqn: Richardson-Dushman}:
\begin{gather}
    J_t=A_R^*T_w^2\exp\left[{\frac{-q}{k_BT_w}\left(\phi_{WF,e}+\delta H(\delta)\right)}\right] \\
    \delta = \phi_{WF,c}+\phi_d-\phi_{WF,e} \nonumber
\end{gather}
where $\phi_d$ is the device voltage, $H$ is the Heaviside function, and $\phi_{WF,e}$ and $\phi_{WF,c}$ are the emitter and collector work functions. A plasma-filled gap may also introduce Schottky enhancement effects and adverse current leakage. When $\phi_{WF,c} = \phi_{WF,a}$, the expression is simplified, and the device voltage can be added to the work function. 
\begin{equation}
    J_t=A_R^*T_w^2\exp\left[\frac{-q}{k_BT_w}\left(\phi_{WF,e}+\phi_d\right)\right]
\end{equation}
The voltage by the TEC can then be determined by setting the net thermionic current equal to that through the device.  

However, even if the plasma sheath is collisionless for ideal ETC systems, the electrode ``gap'' is large. Thermionic electrons thermalize within the flowfield, and a substantial drop in temperature is observed as one travels further from the stagnation point. Not only do these lower temperatures reduce the ionization fraction, but they also decrease the average kinetic energy of the remaining free electrons. When these phenomena are considered together, the result is that electron collection aft the leading edge relies on a smaller population of electrons with a lower average thermal energy overcoming a larger sheath potential barrier. While these disadvantages can be mitigated somewhat by utilizing large acreage regions of the aircraft for electron collection, the assumption that all emitted electrons are collected downstream requires further investigation. This is particularly true for cases with very large plasma densities, as overcompensation is likely to increase both the permitted thermionic current and flowfield electron heating of the wall. Results presented later in this work indicate that preferential collection of electrons at the leading edge could cause diminishing returns when the thermionic space-charge is overcompensated, and can even result in previously unconsidered adverse heating effects when the plasma density is sufficiently large.

\section{Electron Energetics} \label{Sec: ETC energetics}

A fundamental issue arises when one considers Eq.~\ref{Eqn: Basic thermionic electron cooling} in the context of electron collection surface heating. Thermionically emitted electrons alone cannot adequately describe energy transport for a plasma-facing electrode, as heating is not possible without including additional terms. This section discusses the energetics associated with the various charged-particle flows through ETC systems. Some comparisons are made to controlled fusion due to its relatively similar environment.

\subsection{Electron contributions to ETC energy flux}

As suggested by its name, the primary energy carrier in ETC systems are electrons themselves. Each emitted electron must have sufficient energy to overcome the work function at the surface, at which point they are assumed to cool the surface by an amount proportional to the work function plus residual kinetic energy. Common sources of energy for electron emission processes are thermal energy (thermionic emission), photons (photoemission), and kinetic energy of incident particles on the surface (secondary electron emission). The role these emission modes play in ETC are now discussed.  

\paragraph{Thermionic emission}

Thermionic flux from the wall, and the cooling per emitted electron, can be derived using Landauer transport theory and by approximating the Fermi-Dirac electron population with Maxwellian statistics (see supplemental information). 
Cooling occurs only if the emitted electron is replaced by a conduction electron with an expected energy equal to the electrochemical potential.
For simplicity, each emitted electron is assumed to contribute the maximum cooling of $q\phi_{WF} + 2k_BT_w$. If the replacement electron originates from the plasma, the corresponding heating is accounted for separately via a flowfield electron term.

\paragraph{Photoemission}

Photoemission is generally negligible in controlled fusion due to the relatively small photon energy flux at the wall relative to that for charged particles \cite{post2013physics}. Similar logic is employed for ETC, except with a comparison between convective and radiative heating. Radiative heating from the shock becomes prominent for large shock volumes when the temperature reaches about 10,000 K \cite{anderson2019hypersonic}. It is shown later in this work that passive ETC is not practical for the blunt reentry vehicles typically associated with non-negligible radiative heating. It is reasonable to assume photoemission is negligible unless future ETC applications are developed for blunt reentry applications where radiative heating is significant.

\paragraph{Secondary electron emission} \label{Sec: secondary electron emission}

Secondary electron emission (SEE) describes the process through which energy carried by incident particles ejects electrons from the surface. SEE is generally described using secondary electron emission yield, $\delta_{SEE}$, which is expected electron emission per incident particle.

Tolias \cite{tolias2014secondary} calculated and graphed the secondary electron emission yield from a tungsten wall for primary electron energies from 0-2 keV using the Sternglass \cite{sternglass1954theory} and Young-Dekker \cite{lye1957theory, young1957dissipation} formulas. Each incident flowfield electron has an average kinetic energy of $2k_BT_{fe0}$ as it exits the plasma sheath (see supplemental information). All but the most extreme trajectories have primary electron energies of less than 2 eV and negligible $\delta_{SEE}$. 

SEE due to incident ion kinetic energy is generally neglected for fusion plasmas, as the required energy ($\ge$1 keV) is so large that it would also cause unacceptable heating and sputtering of the containing wall \cite{post2013physics}. Hypersonic plasmas are much colder than those used for fusion, so kinetic energy-induced SEE may also be neglected.

Ions can also eject secondary electrons through the potential energy released when undergoing ion-electron recombination within the wall. $\delta_{SEE}$ due to ion-electron recombination \emph{increases} slightly with decreasing kinetic energy \cite{delaunay1987electron, PhysRev.104.672}. For a given velocity, the yield grows linearly with the potential energy at a rate of about 0.01 to 0.02 electrons per eV \cite{post2013physics, delaunay1987electron}. Plasma-seeded ETC is expected to utilize cesium due to its very low first ionization energy of under 4 eV, and so any electron emission resulting from ion-electron recombination is expected to be small. This assumption is supported by empirical data for Cs$^+$ bombardment of a W surface, with a measured $\delta_{SEE}$ of $\lesssim0.01$ for accelerating potentials $\le$100 volts \cite{hyatt1928secondary}. Dominant ions for unseeded hypersonic plasmas have first ionization energies generally ranging from 2-4 times greater than that for Cs. While still small, $\delta_{SEE}$ may not be negligible depending on the exact conditions. In this work, secondary electron emission from ion-electron recombination is assumed to be negligible.  

\paragraph{Flowfield electron bombardment}

While secondary electron emission yields due to flowfield electron bombardment are negligible, the energy carried by impinging flowfield electrons is not. Each incident flowfield electron has an average kinetic energy of $2k_BT_{p}$ as it exits the plasma sheath, but prior to thermalization within the wall, the electron must first pass through a surface region associated with the work function. This region's length is comparable to the separation between atoms in the lattice ($\sim$\AA) while the collisionless plasma sheath is comparable to the Debye length ($\sim$\textmu m). Assuming that the work function region is also collisionless, incoming flowfield electrons gain kinetic energy equal to $q\phi_{WF}$ before thermalizing. The energy flux at the wall per impinging electron can be expressed as:
\begin{equation}
    Q_{fe}=q\phi_{WF}+2k_BT_{p}
\end{equation}

\subsection{Ion bombardment} \label{Subsec: Ion Bombardment}

Ions enter the sheath at the Bohm velocity, $u_B$, before being accelerated by the plasma sheath towards the wall. Just as with the flowfield electrons, the ions must now pass through the work function region at the surface. Again approximating this region as very thin and collisionless, the ions' kinetic energy is \emph{reduced} by $q\phi_{WF}$ prior to thermalizing with the wall. Upon reaching the wall, the ion is neutralized by an electron, which deposits additional energy equal to the ionization potential, $I_0$. The resulting energy deposition at the wall per impinging ion is thus:
\begin{equation} \label{Eqn: ion energetics}
    Q_{i} = \frac{1}{2}m_iu_B^2+q\left(\phi_{w}-\phi_{WF}\right)+I_0
\end{equation}

Similarities to controlled fusion are again noted, as this same equation can be derived from ion backscattering equations utilized for tokamak reactor walls \cite{post2013physics} (see supplemental information).

\subsection{Joule heating}

ETC systems require a conductive wall to replace thermionic electrons at the leading edge. The electronic current through the vehicle structure is accompanied by an additional joule heating term:
\begin{equation}
    Q_j = I_w^2R_w
\end{equation}
Previous ETC studies have found that Joule heating is likely small relative to thermionic cooling \cite{uribarri2015electron, kolychev2022estimation}, but it is included in the present work for comparison with all ETC heat flux terms. 

\subsection{Total ETC energy flux}

Including all terms identified in this section results in an updated expression for the cumulative ETC heat flux:
\begin{equation} \label{Eqn: Actual ETC cooling}
    \begin{split}
    Q_{ETC}=&J_{fe}\left[q\phi_{WF}+2k_BT_{p}\right] \\
    -&J_i\left[q\phi_{WF}-\left(q\phi_w+\frac{m_iu_B^2}{2}+I_0\right)\right] \\
    -&J_t\left[q\phi_{WF}+2 k_BT_w\right] \\
    +&Q_j
    \end{split}
\end{equation}
where negative values indicate cooling. The contributions of each species are summarized in Table \ref{Table: general energy transfer contributions}.
\begin{table}
    \centering
      \caption[Breakdown of ETC energy flux per incident or emitted particle]{Breakdown of ETC energy flux per incident or emitted particle. Joule heating from conduction electrons is also included as a generic $Q_j$ term.}
    \label{Table: general energy transfer contributions}
    \renewcommand{\arraystretch}{1.5}
    \begin{tabular}{|c|c|c|c|} \hline 
         \textbf{Species} & \textbf{Work function} & \textbf{Kinetic Energy} & \textbf{Misc.} \\ \hline 
         Thermionic Electrons & $-q\phi_{WF}$ & $-2k_BT_w$ & \\   \hline 
         Flowfield Electrons & $q\phi_{WF}$ & $2k_BT_{fe0}$ &   \\  \hline
         Ions & $-q\phi_{WF}$ & $q\phi_w+\frac{m_iu_B^2}{2}$ & $I_0$  \\ \hline 
        Conduction Electrons& & & $Q_j$ \\ \hline
    \end{tabular}
\end{table}
Similar expressions have been utilized in hollow cathode applications \cite{cassady2004lithium}, though without joule heating. Within the last few months, another ETC-related work reached an identical expression to that derived here \cite{monroe2025circuit}.

\section{Plasma sheath model} \label{Subsec: Plasma Model}

To determine the heat flux distribution along the vehicle surface, it is first necessary to calculate the charged particle flux as a function of the local wall conditions and that of the adjacent plasma. Assuming that the sheath is not space-charge-limited, the flowfield electron and ion flux to the wall are calculated with Eqs.~\ref{eqn: flowfield electron flux} and \ref{eqn: ion flux} respectively. The thermionic emission current is calculated using the Richardson-Dushman equation (Eq.~\ref{Eqn: Richardson-Dushman}) with a Schottky enhancement from the electric field at the wall (Eq.~\ref{Eqn: Schottky}). $J_t$ and $J_R$ represent thermionic emission flux with and without Schottky enhancement. The ion number density at the sheath edge ($n_{i0}$), flowfield electron temperature at the sheath edge ($T_{fe0}$), wall temperature ($T_w$), work function of the surface ($\phi_{WF}$), and dominant ion species ($m_i$) are assumed to be known.
The electron number density at the sheath edge ($n_{fe0}$), Bohm velocity ($u_B$) and the electric field at the wall ($E_w$) are determined by utilizing the 1D Poisson's equation, the generalized Bohm criterion, and an assumption that the plasma is quasineutral at the sheath edge where $\phi = 0$. The resulting equations are listed below, with the full derivation shown in the supplemental information.
\begin{align}
    \frac{1}{2}\xi^2 
    &= \sqrt{2}\Gamma_t\left[(\Theta_w+\eta-\eta_w)^{1/2}-(\Theta_w-\eta_w)^{1/2}\right] \nonumber \\
    &\quad + e^\eta - 1 
    + \mathcal{N}_{i0}C_B\left[(C_B^2 - 2\eta)^{1/2} - C_B\right] \label{Eqn: Electric Field cold ions} \\
    C_B^2 
    &\ge \frac{1+\frac{\sqrt{2}}{2}\Gamma_t\left(\Theta_w - \eta_w\right)^{-1/2}}{1 - \frac{\sqrt{2}}{4}\Gamma_t\left(\Theta_w - \eta_w\right)^{-3/2}} \label{Eqn: Bohm cold}\\
    \mathcal{N}_{i0} 
    &= \frac{\sqrt{2}}{2}\Gamma_t\left[\Theta_w + \eta_w\right]^{-1/2} + 1 \label{Eqn: Quasineutral no reflection}
\end{align}
where:
\begin{equation}
    \begin{aligned}
        \xi^2 &= \frac{\varepsilon E^2}{n_{fe0}k_BT_{fe0}} \\
        \eta &= \frac{q\phi}{k_B T_{fe0}} \\
        \eta_w &= \frac{q\phi_w}{k_B T_{e0}} \\
        \Theta_w &= \frac{T_w}{T_{fe0}} \\
        C_B &= u_B\sqrt{\frac{m_i}{k_B T_{fe0}}} \\
        \mathcal{N}_{i0} &= \frac{n_{i0}}{n_{e0}} \\
        \Gamma_t &= \frac{J_t}{n_{fe0}}\sqrt{\frac{m_e}{k_B T_{fe0}}} \\
    \end{aligned}
\end{equation}
The electric field at the wall can then be calculated as $\xi_w=\xi(\eta=\eta_w)$. The present model assumes isothermal conditions within the sheath and that the wall is fully catalytic to incoming ions, which are assumed to be cold.

\subsection{Floating potential} \label{Subsec: Floating potential}

Comparison between the actual wall potential, $\phi_w$, to a floating potential, $\phi_f$, gives qualitative insight regarding the direction and magnitude of the electron current between the vehicle surface and adjacent plasma. As the name suggests, the floating potential is the wall potential that would yield zero net local current between the plasma and the wall (Eq.~\ref{Eqn: floating condition}).
By substituting Eqs.~\ref{eqn: flowfield electron flux} and \ref{eqn: ion flux} for $J_{fe}$ and $J_i$ respectively and solving for $\phi_w$, the floating potential can be represented as:
\begin{equation} \label{Eqn: Floating potential}
    \phi_f=\frac{k_BT_{fe0}}{q}\left[\frac{1}{2}\ln{\left(\frac{2\pi m_e}{k_BT_{fe0}}\right)}+\ln{\left(n_{i0}u_B+J_t\right)}-\ln{\left(n_{fe0}\right)}\right]
\end{equation}
$u_B$, $n_{fe0}$, and $J_t$ are functions of $\phi_f$, so an iterative solution is required. For a non-emitting wall, $J_t = 0$ and $n_{fe0} = n_{i0}$, and the Bohm velocity is replaced with the cold-ion acoustic velocity:
\begin{equation} \label{Eqn: cold-ion acoustic}
    u_B = \sqrt{\frac{k_BT_{fe0}}{m_i}}
\end{equation}
The floating potential without emission, $\phi_{f,ne}$, becomes:
\begin{equation} \label{Eqn: Cold ion floating no emission}
    \phi_{f,ne}=\frac{k_BT_{fe0}}{q}\ln{\left[\sqrt{\left(\frac{2\pi m_e}{m_i}\right)}\right]}
\end{equation} 

The actual wall potential is \emph{not} necessarily the same as the floating potential, as an active or passive biasing of the surface permits a net current. If $\phi_w$ is less than $\phi_f$, the flowfield electron flux decreases because the potential barrier shielding the wall from flowfield electrons is larger. The thermionic flux would also slightly increase due to a stronger Schottky enhancement. The result is that there would now be a net emission of electrons out of the wall into the adjacent plasma. Similarly, a $\phi_w$ greater than $\phi_f$ would result in a net accumulation of electrons. This approach is later used to interpret which regions of the vehicle are acting as emitters or collectors in the ETC circuit. The floating potential for a non-emitting wall, $\phi_{f,ne}$, is also useful for qualitative interpretation of emitting surfaces, as the local emission strength relative to the plasma density can be visualized through the difference between $\phi_{f}$ and $\phi_{f,ne}$.

\subsection{Critical potential}

The present sheath model is constructed with an assumption that ion and thermionic electron flux within the sheath is constant, which would not be true should a virtual cathode retard the flow. A virtual cathode forms for wall potentials above the critical potential, which is identified by setting $\xi_w=0$ and solving for the critical wall potential, $\eta_c = \eta_w(\xi_w = 0)$. When solving for $\eta_c$, $J_t$ is equal to the nominal Richardson-Dushman thermionic current without Schottky enhancement, $J_R = J_t(\phi_{WF,\text{eff}}=\phi_{WF})$. For large emission currents, Eq.~\ref{Eqn: Electric Field cold ions} may not return a valid critical potential. For cold-ion sheaths, this situation is avoided by first calculating the critical Richardson emission current, $J_{c}$, by solving Eq.~\ref{Eqn: Electric Field cold ions} for $J_t$ at $\eta_w(\xi_w = 0)$ and taking the limit as $\phi_c$ approaches negative infinity:
\begin{equation} \label{Eqn: critical current}
    J_{c}=n_{i0}u_{B}\sqrt{\frac{m_i}{m_e}}
\end{equation}
If the Richardson-Dushman current exceeds $J_{c}$, a virtual cathode forms regardless of the wall potential and can only be prevented by increasing the plasma density. 

If $\phi_w$ is greater than $\phi_c$, a virtual cathode forms and the plasma sheath model must be revisited. This work aims to study the upper limits of ETC performance without external biasing, so conditions inducing a virtual cathode are not considered. At wall potentials below $\phi_c$, the electric field at the wall enhances thermionic emission.  

\section{Charge emission and accumulation along hypersonic vehicle} \label{Subsec:leading edge model}

\begin{figure}
    \centering
    \includegraphics[width=0.6\textwidth]{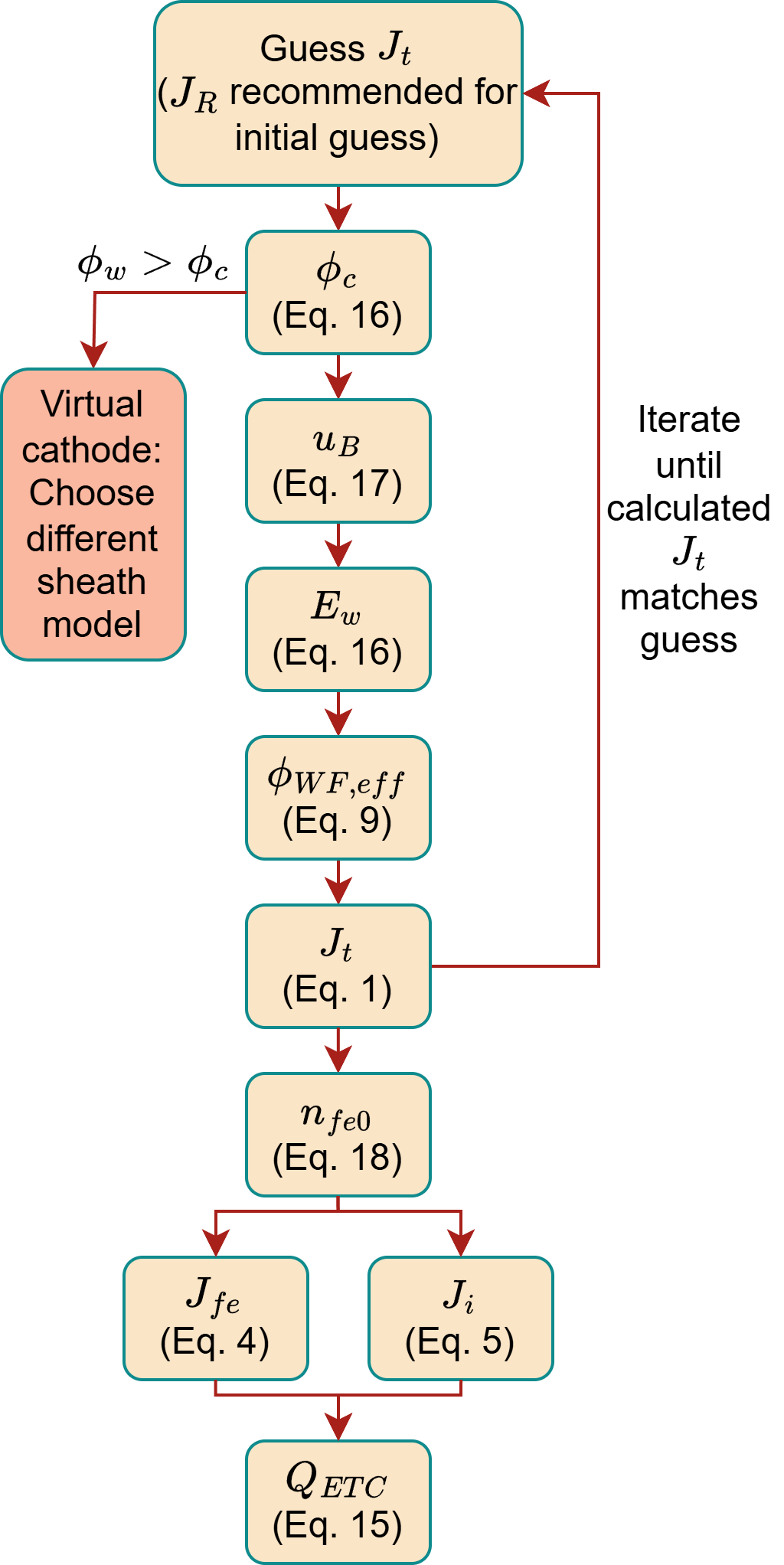}
    \caption[Methodology for calculating ETC heat flux from cold ion sheath model]{\label{fig: Cold ion logic} Methodology utilized to find $Q_{ETC}$ for the cold ion sheath model. Joule heating, $Q_j$, is neglected until the system current through the vehicle surface is determined.
    }
\end{figure}

The sheath model described in the previous section is now incorporated into a discretized model of a hypersonic leading edge to determine the heat flux distribution of ETC along a vehicle surface. At each location along the vehicle surface, the local electron emission or accumulation is calculated from known plasma ($n_{i0}$, $T_{fe0}$, $m_i$) and wall ($T_w$, $\phi_{WF}$, $A^*_R$, $\phi_w$) conditions following the methodology described in Fig.~\ref{fig: Cold ion logic}.
Plasma properties can be determined through CFD solvers \cite{parent2021plasma}, which when coupled with thermal models for the vehicle surface can be used to approximate an initial guess for the temperature distribution along the vehicle surface. An electrical potential guess is then assigned at the node corresponding to the stagnation point, after which the local electron current is calculated. Charge conservation applied within the node dictates that any net electron emission into the plasma at the leading edge must be replaced by conduction from the adjacent downstream node. Electrical conduction is associated with a potential change through Ohm's law, so the potential at each node within the vehicle surface is:
\begin{gather}
    \phi_{n+1} = \phi_n+\frac{\rho_n(\Delta s)^2}{t}\sum_{m=0}^n J_{net,m}
\end{gather}
where $t$ and $\rho_n$ are the hull thickness and resistivity, and $\Delta s$ is the node step size.  
This procedure is continued to successive nodes until the end of the vehicle is reached, at which point the validity of the initial potential guess is checked through global charge conservation. If the initial stagnation potential is too high, there is a net accumulation of electrons and the vehicle becomes more negatively charged. Similarly, a guess that is too low induces net electron emission and increases the overall vehicle potential. Iteration continues until the total cumulative charge exchange between the vehicle and flowfield plasma is zero, at which point the resulting ETC heat flux is calculated via Eq.~\ref{Eqn: Actual ETC cooling} and the temperature distribution can be updated. This process continues until a self-consistent solution is reached.  

\begin{figure}
    \centering
    \includegraphics[width=0.6\textwidth]{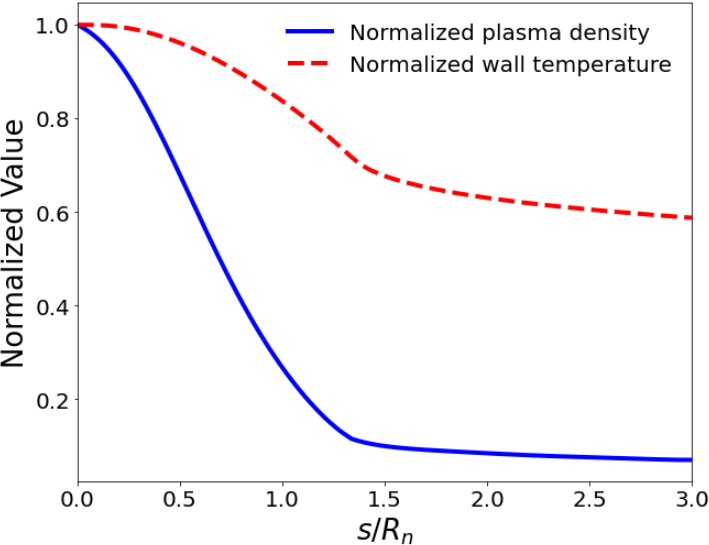}
    \caption[Wall temperature and sheath edge plasma density distributions used in simulations]{\label{fig:Nominal T n distribution} Wall temperature, $T_w$, and sheath edge plasma density, $n_{i0}$. Values are normalized by the values at the stagnation point, which are 2015 K and $3.24\times10^{13} \text{m}^{-3}$ respectively. To prevent the formation of a virtual cathode, $n_{i0}$ is multiplied by a scaling factor $Y$ to obtain values used in simulations.
    }
\end{figure}

The assumed airframe consists of a 3 mm-thick tungsten sheet formed into a wedge with a leading radius of 1 cm. The nominal length of the ETC circuit is 1 m. Tungsten resistivity, $\rho_W$, is assumed to be a function of the average wall temperature between adjacent nodes\cite{desai1984electrical}. The nominal wall temperature and sheath edge plasma density (Fig.~\ref{fig:Nominal T n distribution}) are taken from CFD results by Hanquist \cite{hanquist2017modeling} for this particular geometry traveling at 4 km/s at 60 km altitude. For windward coordinates $s>3R_n$, conditions are approximated by those at $s=3R_n$. The flowfield electron temperature at the sheath edge, $T_{fe0}$, is arbitrarily chosen to be 1.5 times the wall temperature. 
The work function, $\phi_{WF}$, is set to 2.5 eV, corresponding to cesiated tungsten with 27\% surface coverage\cite{sahu2022cesium, gyftopoulos1962work}. The Richardson's constant, $A^*_R$, is assumed to be $6\times10^5$ A/m$^2$K$^2$. The air plasma density associated with this trajectory is not sufficient to compensate thermionic space-charge effects, so Cs seeding near the stagnation point is assumed to prevent the formation of a virtual cathode. Plasma seeding is approximated by multiplying the nominal distribution from Fig.~\ref{fig:Nominal T n distribution} by a scaling constant, $Y$. The minimum $Y$ value to prevent the formation of a virtual cathode, $Y_0$, is approximately $2.8 \times 10^6$, which corresponds to a plasma density at the stagnation point of $9.1\times10^{19} \text{m}^{-3}$. $Y_0$ is sufficiently large relative to the air plasma species that only Cs$^+$ ions are considered. 

While conditions are selected to be representative of hypersonic trajectories, they are not derived from detailed vehicle-specific analysis. Rather, they are intended to demonstrate qualitative trends and limitations of passive ETC systems. Steady state temperature distributions are not computed in this work, but rather heat flux distributions along the vehicle surface are compared in a parametric study of various properties for an assumed temperature distribution.

\section{Results and discussion} \label{Sec: Cold ion Results}

A parametric study is performed to determine the impact of various changes to the conditions of the wall, plasma, and geometry. The conditions for each case simulated are summarized in Table \ref{Table: Case Descriptions}, in which case 1 corresponds to nominal conditions. The heat flux is predicted using Eq.~\ref{Eqn: Actual ETC cooling} and compared to Eq.~\ref{Eqn: Basic thermionic electron cooling}, which is commonly employed in ETC literature, both with and without Schottky enhancements. Additionally, heat flux is compared to that for an ideal ETC system comprising a vehicle with infinite length and negligible electrical resistivity. 
\begin{table}
   \caption[Description of simulated cold ion cases]{Summary of cases. Nominal conditions are shown in case 1. Only differences from case 1 are noted for subsequent cases.}
    \label{Table: Case Descriptions}
    \tiny
    \centering
    \resizebox{\textwidth}{!}{%
    \begin{tabular}{|c|c|c|c|c|c|c|c|c|c|} \hline 
         \textbf{ } & \textbf{$T_w$} & \textbf{$T_p$} & \textbf{$Y$} & \textbf{$\phi_{WF}$} & $A$ & \textbf{$R_n$} & \textbf{$L$} & \textbf{$t$} & \textbf{$\rho/\rho_W$} \\ \hline 

         Case 1 & Fig.~\ref{fig:Nominal T n distribution} & 1.5$T_w$ & $Y_0$ & 2.5 eV & 6$\times10^5$ A/m$^2$K$^2$ & 1 cm & 1 m & 3 mm & $1$ \\ \hline

         Case 2 &  &  & 10$Y_0$ &  &  &  &  &  &  \\ \hline

         Case 3 &  &  & 50$Y_0$ &   &  &  &  &  &  \\ \hline

         Case 4 &  &  &  & 3.5 eV &  &  &  &  &  \\ \hline

         Case 5 &  &  &  & 4.5 eV &  &  &  &  &  \\ \hline

         Case 6 &  &  & 50$Y_0$ & 3.5 eV &  &  &  &  &  \\ \hline

         Case 7 &  &  & 50$Y_0$ & 4.5 eV &  &  &  &  &  \\ \hline

         Case 8 &  &  & 10$Y_0$ &  &  & 2 cm &  &  &  \\ \hline

         Case 9 &  &  & 10$Y_0$ &  &  & 5 cm &  &  &  \\ \hline

         Case 10 &  &  & 10$Y_0$ &  &  & 10 cm &  &  &  \\ \hline

         Case 11 &  &  & 10$Y_0$ &  &  & 20 cm &  &  &  \\ \hline

         Case 12 &  &  & 500$Y_0$ &  &  & 2 cm &  &  &  \\ \hline 
                 
         Case 13 &  &  & 500$Y_0$ &  &  & 2 cm &  &  &  \\ \hline

         Case 14 &  &  & 500$Y_0$ &  &  & 5 cm &  &  &  \\ \hline

         Case 15 &  &  & 500$Y_0$ &  &  & 10 cm &  &  &  \\ \hline

         Case 16 &  &  & 500$Y_0$ &  &  & 20 cm &  &  &  \\ \hline

         Case 17 &  &  & 10$Y_0$ &  &  &  & 0.1 m &  &  \\ \hline

         Case 18 &  &  & 10$Y_0$ &  &  &  & 0.5 m &  &  \\ \hline

         Case 19 &  &  & 10$Y_0$ &  &  &  & 10 m &  &  \\ \hline

         Case 20 &  &  & 10$Y_0$ &  &  &  &  &  & $100$ \\ \hline

         Case 21 &  &  & 10$Y_0$ &  &  &  &  &  & $0.01$ \\ \hline

         Case 22 &  &  & 10$Y_0$ &  &  &  &  &  & $1\times10^{-4}$ \\ \hline

         Case 23 &  &  & 10$Y_0$ &  &  &  & 10 m &  & $1\times10^{-4}$ \\ \hline

         Case 24 &  & $T_p$ &  &  &  &  &  &  &  \\ \hline

         Case 25 &  & $2T_p$ &  &  &  &  &  &  &  \\ \hline

         Case 26 &  & $3T_p$ &  &  &  &  &  &  &  \\ \hline

         Case 27 &  & $T_p$ & 50$Y_0$ &  &  &  &  &  &  \\ \hline

         Case 28 &  & $2T_p$ & 50$Y_0$ &  &  &  &  &  &  \\ \hline

         Case 29 &  & $3T_p$ & 50$Y_0$ &  &  &  &  &  &  \\ \hline

    \end{tabular}
    }
\end{table}

\begin{figure*}[htb]
    \centering
    \includegraphics[width=1\textwidth]{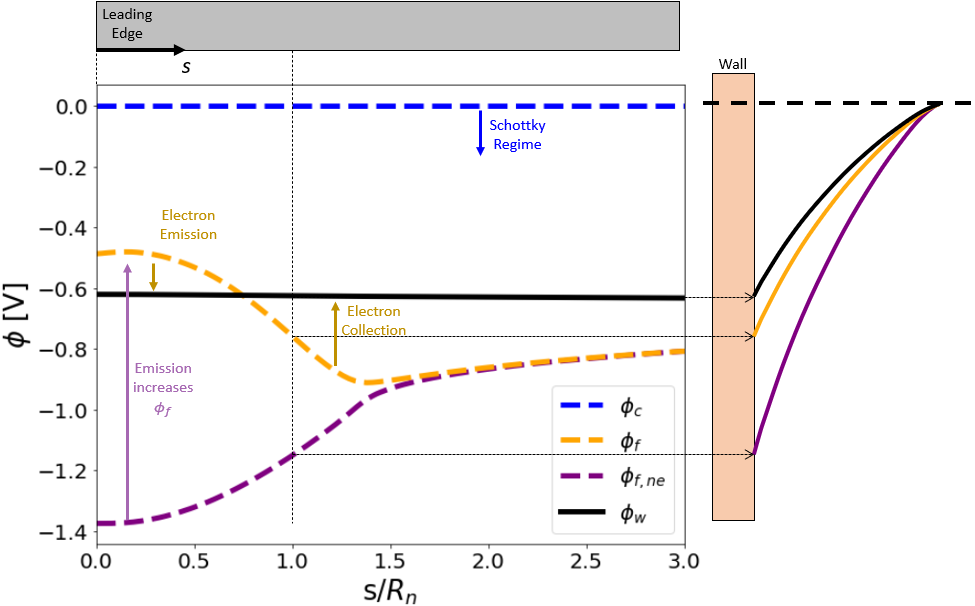}
    \caption[Explanation of results - potential graphs]{Explanation of potential graphs using case 2 as an example. At each location along the vehicle surface $\phi_{f,ne}$, $\phi_{f}$, $\phi_w$, and $\phi_c$ are calculated, from which qualitative insight into local ETC behavior is obtained. $\phi_f$ vs $\phi_{f,ne}$ determines the local emission strength relative to the plasma density. $\phi_f$ can then be compared to $\phi_w$ to ascertain whether the net electron current is into or out of the vehicle. Finally, $\phi_w$ is compared to $\phi_c$ to check whether a virtual cathode forms anywhere along the vehicle surface.}
    \label{fig:Potential graph explanation}
\end{figure*}

Results are presented as a series of graphs showing the energy flux and wall potential along the vehicle surface. The heat flux graphs are relatively straightforward, but additional explanation is merited for the potential graphs. At each discrete location along the vehicle surface, the floating wall potential for a non-emitting wall ($\phi_{f,ne}$), the floating wall potential for an emitting wall ($\phi_f$), and the critical potential ($\phi_c$) are calculated. These are then graphed along with the actual wall potential ($\phi_w$) that satisfies cumulative charge conservation between the vehicle and the plasma. An example for case 2 is shown in Fig.~\ref{fig:Potential graph explanation}. The main qualitative features of these graphs are summarized below:
\begin{itemize}
    \item{$\phi_f \text{ vs.~} \phi_{f,ne}$ - Electron emission from the surface increases the floating potential as described in Sec.~\ref{Subsec: Floating potential}. When emission is small relative to the plasma density, such as colder downstream regions, the two lines converge.}
    \item{$\phi_w \text{ vs.~} \phi_f$ - The wall potential is compared to the floating wall potential to determine the direction and strength of electron emission or accumulation as described in Sec.~\ref{Subsec: Floating potential}. $\phi_w < \phi_f$ indicates local emission,  $\phi_w > \phi_f$ indicates local accumulation. In Fig.~\ref{fig:Potential graph explanation}, there is a net emission of electrons for approximately 7.5 mm, after which accumulation begins. 
    \item{$\phi_w \text{ vs.~} \phi_c$ - So long as the wall potential is less than the critical potential, no virtual cathode forms.}}
\end{itemize}

The concept of an ideal collector with an infinite length and electrical conductivity can now be introduced. The infinite collector length enables the collection of all electrons with $\phi_w \approx \phi_{f,collector}$. As there is no resistance to electron conduction back towards the leading edge, $\Delta\phi_w=0$, and the entire vehicle would have a surface potential equal to the floating condition within the collector region far from the leading edge. This process represents the ideal operation of a passive ETC system and is referred to as the ideal collector model.  

\subsection{Effect of plasma density and overcompensation}

\begin{figure}[htbp]
    \centering
    \includegraphics[width=0.7\textwidth]{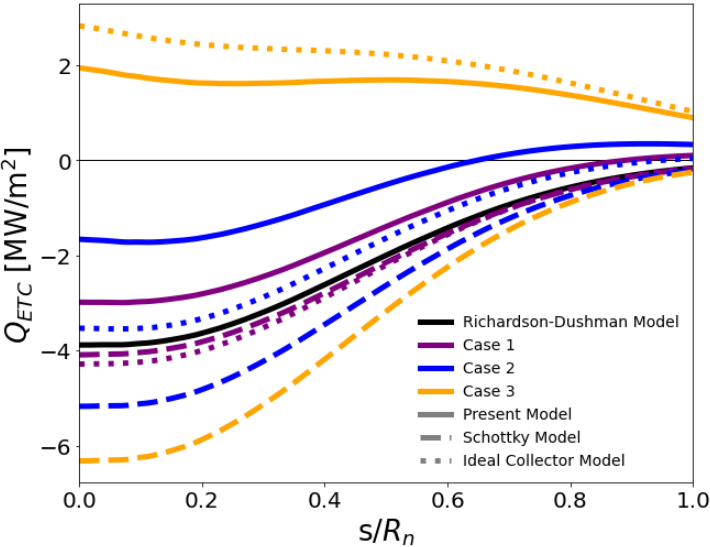}
    \caption[Impact of plasma density and overcompensation on ETC performance]{\label{fig:Plasma density impact} ETC energy flux is graphed for cases 1-3 utilizing the Richardson-Dushman model (Eq.~\ref{Eqn: Basic thermionic electron cooling} without Schottky enhancement), the Schottky model (Eq.~\ref{Eqn: Basic thermionic electron cooling} with Schottky enhancement), the ideal collector model (Eq.~\ref{Eqn: Actual ETC cooling} where $\phi_w = \phi_{f,collector}$) and the present model (Eq.~\ref{Eqn: Actual ETC cooling}).}
\end{figure}

The energy flux as a function of position is shown for several plasma densities in Fig.~\ref{fig:Plasma density impact}. The scaling factor, $Y$, was chosen to be $Y_0$, $10Y_0$, and $50Y_0$, which correspond to stagnation point plasma densities of $9.1\times10^{19} \text{m}^{-3}$, $9.1\times10^{20} \text{m}^{-3}$, and $4.5\times10^{21}$ $\text{m}^{-3}$ respectively. Although a detailed analysis of the Cs injection required to achieve such plasma densities is beyond the scope of this work, it is worth noting that a previous study predicted a plasma density of approximately $10^{21}\text{ m}^{-3}$ for a Cs transpiration rate of $0.8\text{ kg}\text{m}^{-2}\text{s}^{-1}$ and a 1 cm leading edge radius at Mach 14 and an altitude of 60 km\cite{sahu2022full}. 

Nominal thermionic cooling without Shottky enhancement (i.e., Richardson-Dushman Model) stays the same regardless of the surrounding plasma density so long as no virtual cathode forms. However, the Schottky enhanced emission model (Eqs.~\ref{Eqn: Basic thermionic electron cooling} and \ref{Eqn: Schottky}), the present model (Eq.~\ref{Eqn: Actual ETC cooling}), and ideal collector model (Eq.~\ref{Eqn: Actual ETC cooling} with $\phi_w=\phi_f(s/R_n \ge 3)$) all exhibit different behavior as the plasma density changes. 

When the thermionic space-charge is exactly compensated (case 1) the Schottky emission model predicts a stagnation point cooling flux of approximately 4.1 $\frac{\text{MW}}{\text{m}^2}$, which is a marginal improvement over the 3.9 predicted when assuming that emission flux is equal to the nominal Richardson-Dushman current. It is also slightly below the 4.3 $\frac{\text{MW}}{\text{m}^2}$ predicted by the ideal collector model. The present model predicts 3 $\frac{\text{MW}}{\text{m}^2}$ of cooling, which is a 27\% reduction from the Schottky model. However, even with the reduced performance, ETC is still significant for all models considered.  

Issues begin to emerge when the plasma sheath becomes overcompensated. Increasing $Y$ by an order of magnitude (case 2) causes the predicted Schottky model cooling to increase by 26\%, while the present and ideal collector models predict ETC reduction of 44\% and 18\% respectively. Upon increasing the plasma density by an additional factor of 5 (case 3) the present and ideal collector models predict \emph{heating} of the leading edge. At the same time, the Schottky model continues to predict enhanced performance with a 55\% improved stagnation point cooling relative to case 1.  

\begin{figure}
    \centering
    \includegraphics[width=0.7\textwidth]{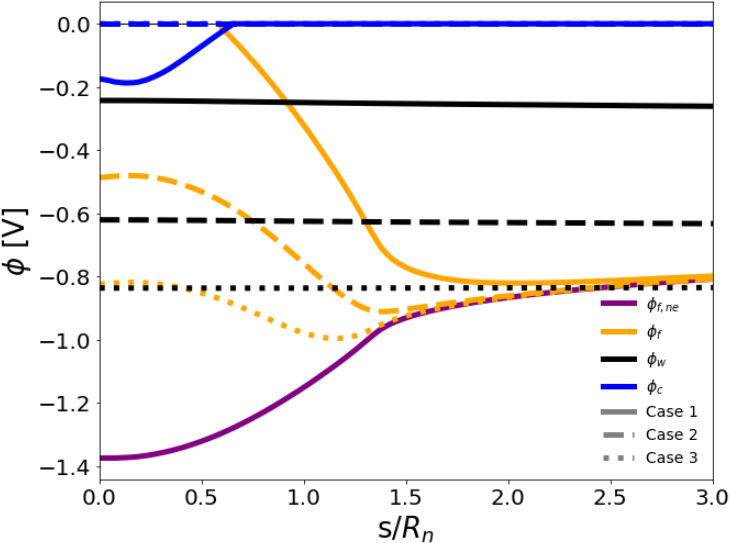}
    \caption[Impact of plasma density and overcompensation on ETC performance - potential graph]{\label{fig:Potential graph for plasma density} The wall potential along the vehicle surface ($\phi_w$) is compared to floating potential with and without emission ($\phi_f$ and $\phi_{f,ne}$ respectively) as well the critical potential for cases 1-3.
    }
\end{figure}

The trends in heat flux versus plasma density are explained using Fig.~\ref{fig:Potential graph for plasma density}. 
The comparison between $\phi_w$ and $\phi_f$ is particularly important. The following must be true for a passively biased ETC TPS to achieve cooling at the leading edge:
\begin{enumerate}
    \item{$\phi_w < \phi_f$ near the stagnation point to have net electron emission}
    \item{$\phi_w > \phi_f$ downstream to have net electron accumulation}
    \item{$\frac{d\phi_w}{ds} < 0$ to conduct electrons from the collector region to the emitter region}
\end{enumerate}
When considered together, these constraints require $\phi_{f,LE} > \phi_{f,collector}$. The larger this difference, the better the ability for the ETC circuit to pump collected electrons from aftbody regions to the stagnation point. 

The floating potential for a non-emitting wall, $\phi_{f,ne}$ can be considered a starting condition. Downstream, the wall temperature is small enough that thermionic emission is negligible and $\phi_f \approx \phi_{f,ne}$. It is only upstream where the temperature is large enough to significantly alter $\phi_f$. It is for this reason that the temperature and plasma density distribution at large windward coordinates were chosen to match those at $s/R_n = 3$ to maximize ETC performance. At larger coordinates the plasma temperature decreases, thereby increasing $\phi_{f,ne}$ per Eq.~\ref{Eqn: Cold ion floating no emission} and inhibiting electron collection downstream. This phenomenon matches earlier qualitative predictions that electrons in a colder, less dense plasma would have more difficulty penetrating the plasma sheath. Keeping plasma and wall properties constant for $s/R_n > 3$ leads to larger ETC performance for the present sheath model, which is a reasonable approximation when trying to study the limitations of passive systems under the most advantageous set of assumptions.  

When the thermionic space-charge is exactly compensated, $\phi_f$ is much larger than $\phi_{f,ne}$, and the circuit can easily conduct electrons to the leading edge. As $Y$ increases, similar thermionic fluxes create a smaller change from $\phi_f \text{ to } \phi_{f,ne}$ as flowfield electrons are more plentiful. In the presence of significant overcompensation, $\phi_f$ at the leading edge is no longer more positive than $\phi_{f,ne}$ downstream and the ETC circuit ceases to function as intended. If the overcompensation increases further, electron conduction reverses direction and heats the leading edge. The ideal collector model predicts even greater heating, as an infinite conductivity does nothing to retard this adverse conduction.  

\subsection{Effect of work function}

\begin{figure}[htb]
  \centering
  \begin{minipage}{0.49\textwidth}
    \includegraphics[width=\linewidth]{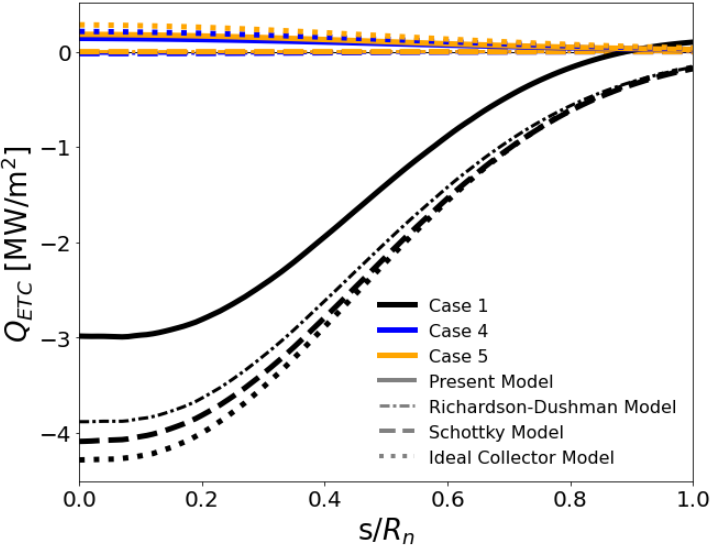}
    \caption*{a) Cases 1, 4, 5}  
  \end{minipage}
  \hfill
  \begin{minipage}{0.49\textwidth}
    \includegraphics[width=\linewidth]{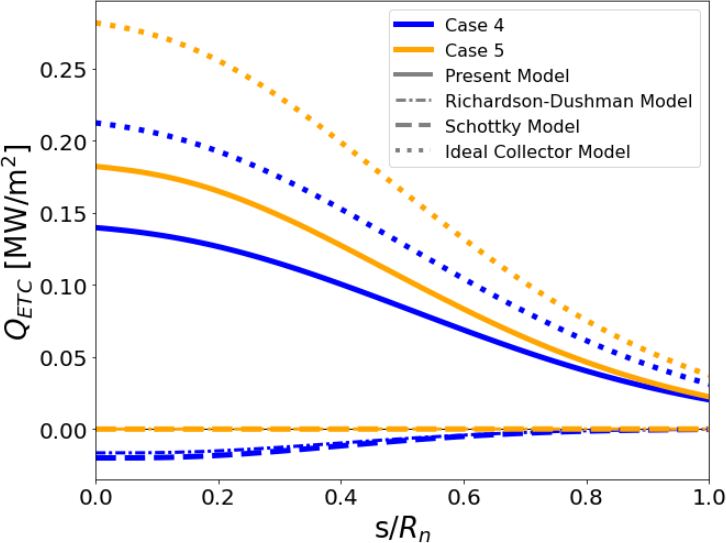}
    \caption*{b) Cases 4, 5}  
  \end{minipage}
  \caption[Impact of work function on ETC performance]{Surface heat flux for nominal plasma density with varying $\phi_{WF}$. a) Shows cases 1 ($\phi_{WF} = 2.5$ V), 4 ($\phi_{WF} = 3.5$ V), and 5 ($\phi_{WF} = 4.5$ V). b) Shows cases 4 and 5 only for improved visual clarity.}
  \label{Fig: Work function impact 1}
\end{figure}

\begin{figure}
    \centering
    \includegraphics[width=0.7\textwidth]{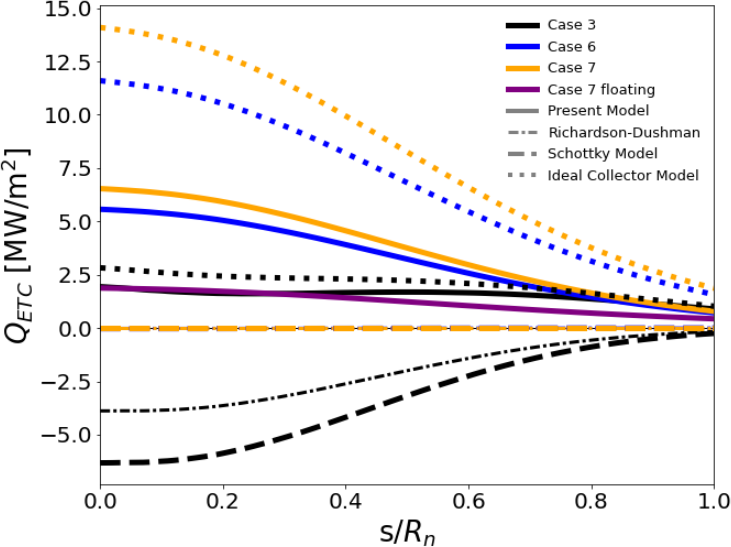}
    \caption[Impact of work function on ETC performance at higher plasma density]{\label{fig:Work function impact 2} Surface heat flux for $Y=50Y_0$ with varying $\phi_{WF}$. Cases 3, 6, and 7 correspond to $\phi_{WF}$ of 2.5, 3.5, and 4.5 V respectively. Case 7 is also shown with an alternative floating case, where $\phi_w = \phi_f$ for all locations along the vehicle.  
    }
    \label{Fig: Work function impact 2}
\end{figure}

ETC performance depends on the thermionic emission being sufficiently large relative to the plasma density, so increasing $\phi_{WF}$ should produce a result similar to that for increasing $n_{i0}$. Figs. \ref{Fig: Work function impact 1} and \ref{Fig: Work function impact 2} show energy flux distributions for varying $\phi_{WF}$ at $Y=Y_0$ and $50Y_0$ respectively.

For both plasma densities a work function of 3.5 V was sufficient to effectively deactivate thermionic emission and induce surface heating. As the work function increases further, surface heating increases as a direct result of the additional kinetic energy imparted to flowfield electrons by the work function surface potential. While the expected stagnation heating for the $Y=50Y_0$ case is extremely large, it is unlikely that such a sharp leading edge radius would be chosen for a trajectory capable of producing such large plasma densities. If a sharp leading edge \emph{were} required, a dielectric surface could be utilized to mitigate adverse electron conduction. Fig.~\ref{fig:Work function impact 2} shows the impact of such a coating through "Case 7 floating", for which $\phi_w = \phi_f$ at all locations along the surface. While surface heating is not completely eliminated, there is a 71\% reduction from the standard Case 7 conditions.  

\subsection{Effect of leading edge radius}

Rather than utilizing a sharp leading edge with a dielectric coating, a blunt leading edge is more practical for trajectories with large plasma densities. It was claimed in Sec.~\ref{Sec: Basic theory} that blunt vehicles with larger leading edge radii are expected to behave as floating emitters. While this limits their utility in passive ETC applications, it also means that blunt bodies should be less susceptible to adverse heat pumping effects associated with overcompensated ETC circuits.

The impact of $R_n$ is difficult to isolate, as changes to the vehicle geometry impact the flowfield environment by changing the shock curvature, volume, and temperature. The primary focus here is the shock geometry, as larger $R_n$ reduces bow shock curvature near the stagnation point and reduces spatial variations in the plasma. This effect is studied by keeping the same temperature and plasma density from Fig.~\ref{fig:Nominal T n distribution}, but stretching the distributions by updating $R_n$ for the normalized position. While this abstraction is entirely non-physical, stretching the distributions in this way enables a degree of comparison between a ``blunt'' body exposed to a similar environment as the nominal $R_n$.

\begin{figure}[htbp]
  \centering
  \begin{minipage}{0.49\textwidth}
    \includegraphics[width=\linewidth]{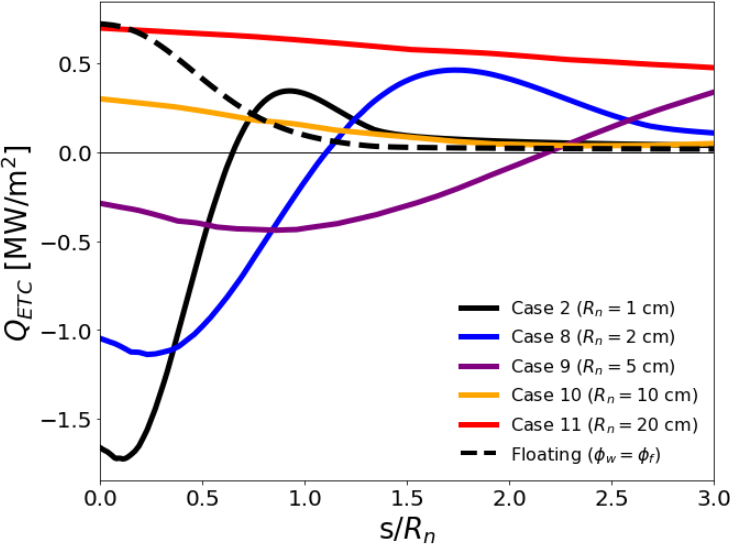}
    \caption*{a) $Y=10Y_0$}  
  \end{minipage}
  \hfill
  \begin{minipage}{0.49\textwidth}
    \includegraphics[width=\linewidth]{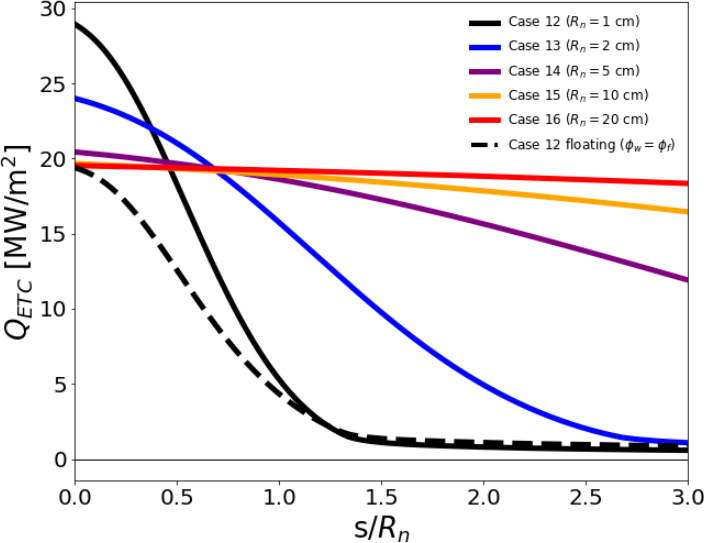}
    \caption*{b) $Y=500Y_0$}  
  \end{minipage}
  \caption[Impact of leading edge radius on ETC performance]{Surface energy flux for varying $R_n$ at a)  $Y=10Y_0$ b) $Y=500Y_0$. A floating condition is also included in each graph for comparison, and is the predicted heating if the $\phi_w = \phi_f$ everywhere along the vehicle surface.}
  \label{Fig: LE radius impact heatflux}
\end{figure}

Fig.~\ref{Fig: LE radius impact heatflux} shows the surface energy flux for varying $R_n$ and $Y$. 
$Y=Y_0$ would have led to the formation of a virtual cathode as $R_n$ was increased, so $Y=10Y_0$ (fig \ref{Fig: LE radius impact heatflux}a) was chosen to demonstrate the impact of $R_n$ for conditions where ETC was initially cooling the surface. As expected, increasing $R_n$ reduces ETC performance and eventually causes surface heating. The energy flux at the stagnation point approaches that for a floating wall as $R_n$ increases. $Y=500Y_0$ (fig \ref{Fig: LE radius impact heatflux}b) was chosen to represent conditions where the flowfield plasma would cause significant surface heating. As $R_n$ is increased, surface heating \emph{decreases} and approaches the stagnation heatflux associated with a floating wall.

\begin{figure}
    \centering
    \includegraphics[width=0.7\textwidth]{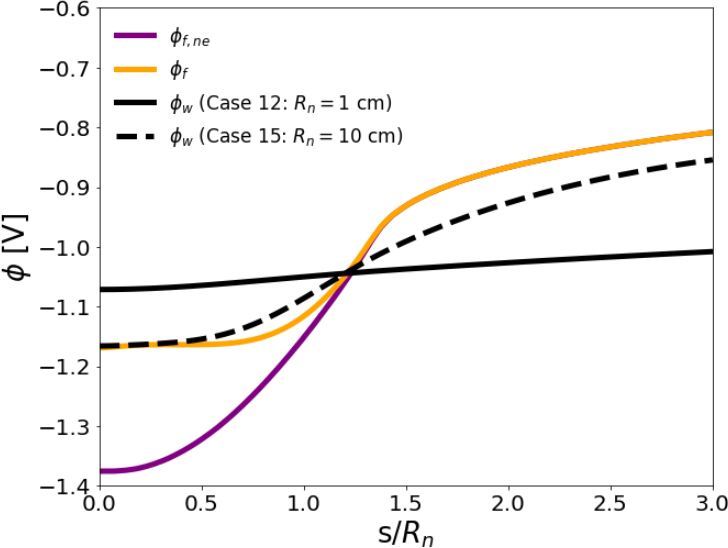}
    \caption[Impact of plasma density and overcompensation on ETC performance - potential graph]{\label{fig: LE radius impact 2} Wall potential distribution for case 12 ($R_n = 1$ cm) and case 15 ($R_n = 10$ cm). Both cases have identical $\phi_f$ and $\phi_{f,ne}$}
\end{figure}

The wall potential distribution for cases 12 ($R_n=1$ cm) and 15 ($R_n=10$ cm) are compared in Fig.~\ref{fig: LE radius impact 2}. As expected, the case with the larger $R_n$ more closely tracks $\phi_f$, especially near the stagnation point.

\subsection{Effect of vehicle length}

\begin{figure}
    \centering
    \includegraphics[width=0.7\textwidth]{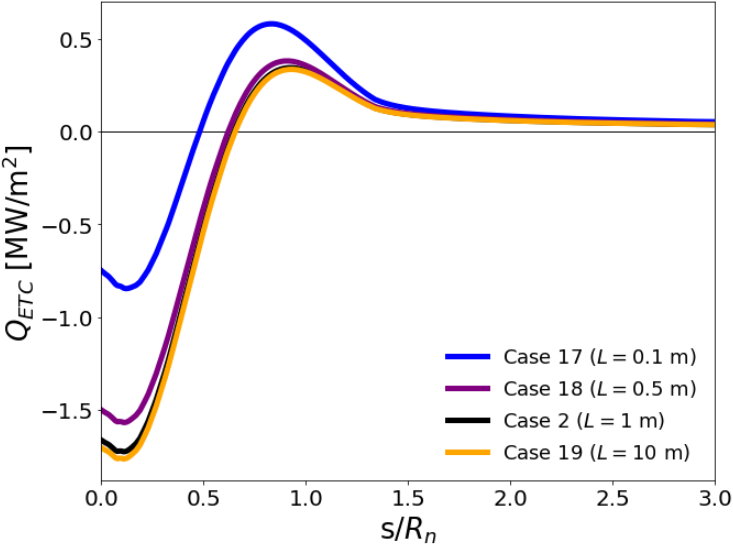}
    \caption[Impact of system length on ETC performance]{\label{fig:Length impact} Surface heating distribution for varying ETC system length at $Y=10Y_0$.
    }
\end{figure}

While electron collection downstream may be more challenging due to the potential distribution and plasma conditions, employing sufficiently large collector areas should mitigate this issue. Fig.~\ref{fig:Length impact} shows energy flux distributions where the length has been varied from 0.1 to 10 m. This length could represent the vehicle geometry or that of an ETC subsystem.  
Increasing the vehicle length above nominal conditions had a negligible impact on ETC performance. Cooling is reduced by 10\% and 55\% for length reductions to 0.5 and  0.1 m respectively.

Even for the longest lengths considered, ETC still performed less than half as well as that for an ideal collector. Increasing the length led to diminishing returns, with moderate ETC cooling still expected for a system even an order of magnitude shorter than the nominal length. 

\subsection{Effect of wall resistance}

\begin{figure}
    \centering
    \includegraphics[width=0.7\textwidth]{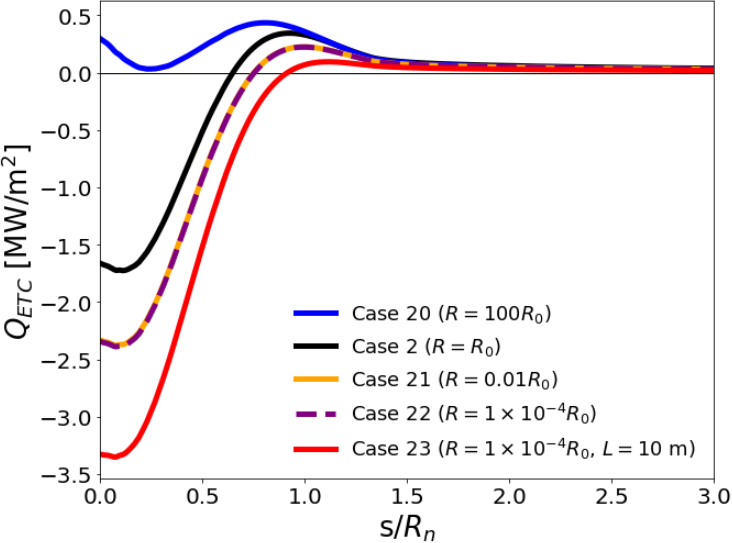}
    \caption[Impact of wall electrical resistance on ETC performance]{\label{fig:Resistance impact} Surface heat flux distribution for varying $\rho$.
    }
\end{figure}

The diminishing returns seen for increasing vehicle lengths implies that the electrical conductivity of the structure is a limiting factor. All conditions tested were derived from case 2, except that the temperature dependent resistivity of the tungsten ($\rho_W$) was scaled by a constant. Results for resistivities ranging from $1\times10^{-4}$ to 100 $\rho_W$ are shown in Fig.~\ref{fig:Resistance impact}.
As expected, increasing the resistance of the ETC circuit signifantly reduces ETC performance. The lower the resistance through the vehicle structure, the greater the ability of the vehicle to recirculate collected electrons. Two orders of magnitude is sufficient to transition from cooling to heating at the leading edge. Lowering the resistance by two orders of magnitude improved ETC by about 40\%, but a further reduction led to negligible changes. When the lowest resistance was combined with an increased vehicle length of 10m, performance was very similar to that of an ideal collector.

Regardless of how long a vehicle may be, the effective collector area is limited by the maximum difference in the floating potential relative to the electrical resistance. As the resistance decreases, larger portions of the vehicle are ``activated'' and additional current can be conducted through the ETC circuit.  

\subsection{Energy flux by species and source}

\begin{figure}[htbp]
  \centering
  \begin{minipage}{0.49\textwidth}
    \includegraphics[width=\linewidth]{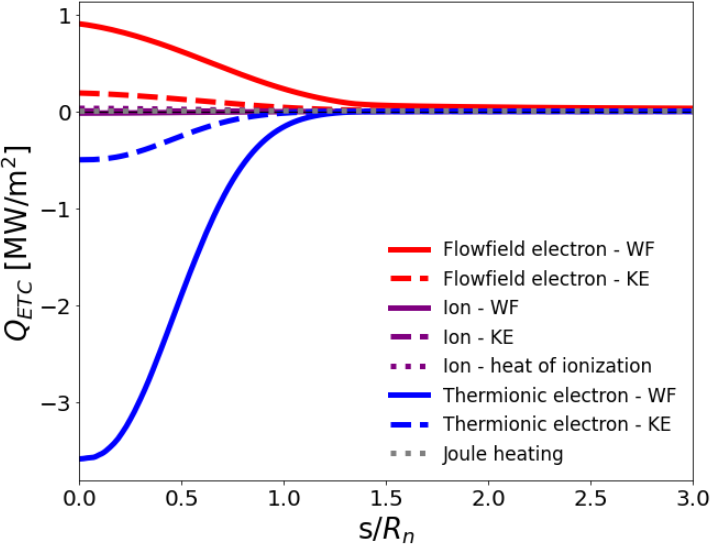}
    \caption*{a) $Y=Y_0$}  
  \end{minipage}
  \hfill
  \begin{minipage}{0.49\textwidth}
    \includegraphics[width=\linewidth]{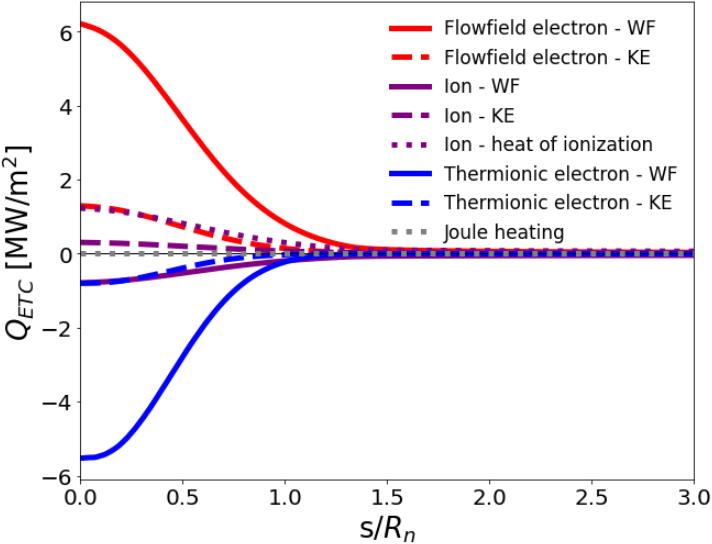}
    \caption*{b) $Y=50Y_0$}  
  \end{minipage}
  \caption[Heat flux breakdown along vehicle surface]{Energy flux breakdown by species and classification for a) $Y=Y_0$ b) $Y=50Y_0$.}
  \label{Fig: Species dependent energy flux}
\end{figure}

Further insight is gained by breaking down surface heat flux into the constituent terms summarized in Table \ref{Table: general energy transfer contributions}. Results for cases 1 and 3 are shown in Fig.~\ref{Fig: Species dependent energy flux} and Table \ref{table: Energy flux breakdown cold ion}.
\begin{table} 
    \caption[Energy flux breakdown at the stagnation point for cases 1 and 3]{Energy flux breakdown [MW/m$^{2}$] at the stagnation point for cases 1 and 3.}
    \label{table: Energy flux breakdown cold ion}
    \centering
    \begin{tabular}{|c|c|c|} \hline 
         \textbf{ } & \textbf{Case 1} & \textbf{Case 3} \\ \hline 
         Thermionic electron, $\phi_{WF}$ & -3.59 & -5.53 \\ \hline 
         Thermionic electron, $KE$ & -0.50 & -0.80 \\ \hline 
         Flowfield electron, $\phi_{WF}$ & 0.90 & 6.22 \\ \hline 
         Flowfield electron, $KE$ & 0.19 & 1.30 \\ \hline 
         Ion, $\phi_{WF}$ & -1.9$\times10^{-2}$ & -0.79 \\ \hline 
         Ion, $KE$ & 3.4$\times10^{-3}$ & 0.31 \\ \hline 
         Ion, recombination & 3.0$\times10^{-2}$ & 1.23 \\ \hline 
         Joule heating & 5.67$\times10^{-7}$ & 7.2$\times10^{-9}$ \\ \hline 
         Total & -2.99 & 1.94 \\ \hline 
    \end{tabular}
\end{table}
Thermionic and flowfield electrons are the most dominant energy carriers between the wall and the plasma sheath. In case 1, where the sheath is minimally compensated, energy fluxes associated with ions are negligible. Ion contributions to energy flux increase with the degree of overcompensation, but are still smaller than that for the electrons. Joule heating is negligible for all cases considered within this study, which supports conclusions within the existing literature \cite{uribarri2015electron, kolychev2022estimation}.

\section{Conclusions}

While passive ETC demonstrates potential, excessive thermionic space-charge can result in surface heating rather than cooling. ETC performance can remain viable within a limited margin of overcompensation, but excessive levels must be avoided. Improperly designed seeded plasma flows can yield plasma densities far exceeding those typical of sharp leading edges. Nevertheless, this study does not discount the viability of plasma seeding, which remains a promising approach when appropriate flow rates are employed for a given trajectory and flowfield environment.

Passive ETC functionality is determined qualitatively through the floating potential distribution along the surface. It is only when $\phi_{f,LE} > \phi_{f,collector}$ that the circuit is capable of pumping electrons through the structure to replace emitted electrons. Highly conductive surfaces enable the ETC circuit to access a greater portion of the vehicle surface, significantly enhancing performance at a fixed potential difference. The best possible performance corresponds to an ideal collector with infinite length and conductivity, but both area and conductivity suffer diminishing returns. Reasonable performance can likely be obtained with standard refractory metals and collector lengths on the order of 10 cm.  

When thermionic emission is insufficient relative to the plasma density, such that $\phi_{f,LE} < \phi_{f,collector}$, adverse electron conduction away from the leading edge exacerbates heating. Flowfield surface heating of sharp leading edges can be mitigated, but not eliminated, through the use of a dielectric coating. Alternatively, a vehicle exposed to large plasma densities can mitigate surface heating through the use of blunt surfaces. Larger leading edge radii are associated with reduced bow shock curvature near the stagnation point, which causes such vehicles to naturally behave as a floating surface in the vicinity of the leading edge. However, this same effect also precludes the use of blunt bodies in passive ETC applications.  

Analysis of energy flux by species and source confirms that thermionic and flowfield electrons dominate ETC energy transfer. Ion contributions to the energy flux is generally small, but may or may not be negligible depending on the conditions. Joule heating remains negligible for all cases considered, justifying its omission from ETC analysis.

\bibliography{main}

\end{document}